\documentclass{mn2e}
\usepackage{natbib}
\usepackage{graphicx}
\usepackage{pslatex}

\begin{document}

\title{The death of FRII radio sources and their connection with radio relics}

\author[C.R. Kaiser \& G. Cotter]{Christian R. Kaiser$^1$\thanks{email: crk@astro.soton.ac.uk} and Garret Cotter$^{2}$\\
$^1$ Department of Physics \& Astronomy, University of Southampton, Southampton SO17 1BJ\\
$^2$ Cavendish Laboratory, Madingley Road, Cambridge, CB3 0HE}

\maketitle

\begin{abstract}
Radio relic sources in galaxy clusters are often described as the
remnants of powerful radio galaxies. Here we develop a model for
the evolution of such relics after the jets cease to supply energy
to the lobes. This includes the treatment of a relic overpressured
with respect to its gaseous surroundings even after the jets
switch off. We also determine the radio emission of relics for a
large variety of assumptions. We take into account the evolution
of the strength of the magnetic field during the phase of
relativistic particle injection into the lobes. The resulting
spectra show mild steepening at around 1\,GHz but avoid any
exponential spectral cut-offs. The model calculations are used to
fit the observed spectra of seven radio relics. The quality of the
fits is excellent for {\it all} models discussed. Unfortunately,
this implies that it is virtually impossible to determine any of
the important source parameters from the observed radio emission
alone.
\end{abstract}


\section{Introduction}

Diffuse radio emission without an apparent host galaxy is found in
an increasing number of clusters of galaxies
\citep[e.g.][]{gtf99,gf00,ks01,gfg01}. These objects can roughly
be divided into radio halos close to the cluster centre which
appear comparatively smooth in radio images and radio relics with
a far more distorted and knotty appearance. The latter are usually
found further away from the central regions \citep{sr84}. Here we
concentrate on radio relic sources.

A large number of possible explanations for radio relics has been
proposed. All of these aim at explaining the production of a
population of relativistic electrons responsible for the observed
radio synchrotron emission. The suggested mechanisms range from
the turbulent wakes of galaxies moving in the cluster's
gravitational potential to cluster mergers. In the present paper
we concentrate on the picture of relics as remnants of once
powerful radio galaxies or radio-loud quasars. This scenario is
based on the idea that the relics are the lobes of former radio
galaxies, the jets of which ceased to supply them with energy some
time ago \citep[e.g.][]{kg94,srm01}. The relativistic electrons,
the emission of which we observe today, were accelerated at the
strong shocks at the ends of the active jets. The electrons lose
their energy due to synchrotron losses, inverse Compton scattering
of cosmic microwave background photons and, possibly, further
adiabatic expansion of the relics themselves. As the relativistic
electrons are not replenished by the active jets, the radiative
energy losses introduce a spectral cut-off moving towards low
frequencies in time. This can potentially explain the observed
steep radio spectra of relics. 

Note however that not all relics may be consistent with the picture of
the passively fading remnant of a radio galaxy. Some relics are
extended and show a rather smooth structure resembling in some ways
the appearance of radio haloes, while others are composed of a number
of individual knots. The variety of morphologies different from the
appearance of active radio galaxies may very well indicate that our
interpretation of relics as the remnants of radio galaxies is
incorrect in many cases. However, it is not clear what the fate of the
material forming the large scale radio structure of a formerly active
radio galaxy is. Buoyant rise in the gravitational potential well of
the cluster will certainly play a role \citep[e.g.][]{cbkbf00}, but
the movement of the parent galaxy through the cluster gas and cluster
mergers may fragment the remnant as well. It is therefore not
surprising that the morphology of relics is seldomly reminiscent of
active radio galaxies. Even if all relics were indeed, as we assume
here, the remnants of radio galaxies, the evolution of the remnant may
imply significant departures from the simple pressure evolution
envisaged in the model introduced here.

Another problem is the possibility of re-acceleration of the
relativistic electrons even after the jets have switched
off. \citet{srm01} present radio observations of the relics in A13,
A85, A133 and A4038. The substantial substructure in these sources may
well be the signature of some highly dynamic processes being at work
which could lead to further acceleration of relativistic particles. In
this paper we will neglect any re-acceleration processes which may
take place in relics.

The question then is whether or not a passively evolving
population of relativistic electrons produces a radio spectrum
consistent with observations. The main problem is that the radio
spectra of relics are shallow at around 100\,MHz and considerably
steeper in the GHz; but they do not show exponential cut-offs. The
spectra are therefore not consistent with a single population of
electrons accelerated all at the same time and passively losing
their energy. Taking this extended injection period into account,
\citet{kg94} showed that two breaks appear in the spectrum. The
lower break is located at the frequency cut-off of the `oldest'
electrons while the higher break is determined by the `youngest'
electrons. The comparatively mild steepening of the spectra
between the breaks is roughly consistent with the observations.
Nevertheless, \citet{kg94} found that for effective pitch-angle
scattering of the relativistic electrons the steepening of the
spectrum was still too strong. Therefore they had to suppress
pitch-angle scattering to further flatten their model spectra.

The exponential cut-off in the spectrum is also avoided in the
case of inhomogeneous magnetic field strengths inside the relic.
The relativistic electrons spend most of their time in the
low-field regions and only occasionally diffuse into the
high-field regions. Thus their energy losses are reduced and by
careful adjustment of the efficiency of the diffusion process an
emission spectrum with mild steepening at high frequencies can be
achieved \citep{pt93,emw97}. \citet{srm01} fitted a model based on
this idea to the spectra of relic sources and found good
agreement.

\citet{kg94} assumed that the relics or their progenitors were in
pressure equilibrium with their surroundings even at the time of
the injection of relativistic electrons. In this paper we expand
this model by taking into account the evolution of the lobes
during the injection of relativistic particles. If the lobes of
radio galaxies of type FRII \citep{fr74} are the progenitors of
the radio relics, then during the time the jets are still active,
the lobes are expanding into the surrounding material. This
implies that the pressure inside the lobes and therefore also the
strength of the magnetic field are not constant during the
particle injection. In fact, the strength of the magnetic field is
decreasing and thus relativistic electrons injected at later times
can survive for longer. We show that this effect results in radio
spectra of the resulting relic sources without exponential
cut-offs and a spectral slope comparable to that in observed
relics.

At the time the jets switch off, the lobes may still be
overpressured with respect to the ambient gas and therefore
continue to expand. We derive expressions for the temporal
behaviour of the volume and the pressure of such `coasting'
relics. We take such a coasting phase into account in our
calculation of model radio spectra from relic sources. Finally, we
show that our models fit the observed spectra very well.
Unfortunately, we find that none of the important source
parameters, like the source age, or the physical processes taking
place in the relics can be determined from the radio spectra
alone.

In Section \ref{sec:spherical} we derive the temporal behaviour of
a coasting relic source after the jets have switched off. We use
the results found to calculate the radio spectra of relic sources
in Section \ref{sec:radio}. Here we also show how the variation of
the main model parameters influences the results. In Section
\ref{sec:compa} we fit the observed spectra of relic sources with
our model and show that very little useful information about the
conditions in the sources and their environments can be gained
from the models. We summarise our results in Section
\ref{sec:conc}.

\section{The spherical radio source}
\label{sec:spherical}

After the jets of a powerful radio sources switch off, the cocoon
inflated during the active phase may still be overpressured with
respect to the external medium. It will then enter the `coasting
phase' while continuing to expand behind a strong bow shock. This
situation is somewhat reminiscent of the early evolution of a
supernova remnant or the explosion of a nuclear weapon. The
dynamics of a strong explosion in a uniform atmosphere are
described by the similarity solution of Taylor and Sedov
\citep[e.g.][]{ls59,ll87}. It is straightforward to generalise
this solution for the case of a power-law density distribution in
the external medium with $\rho = \rho _0 (r/a_0)^{-\beta}$, where
$a_0$ is the scale height or core radius of the distribution. We
will assume this density distribution for all model calculations
in this paper. The radius of the spherical bow shock then grows as
$R_{\rm s} \propto t^{2/(5-\beta)}$. Despite the fact that the
cocoons of radio galaxies are usually not spherical, one might
expect that their expansion during the coasting phase is governed
by the same proportionality once the jet activity has ceased.
However, in the following we will show that the presence of the
contact discontinuity defining the cocoon boundary in radio
galaxies requires a different dynamical behaviour.

\citet{gr94} present a different solution to this problem based on
their assumption that the dynamics of the radio source expansion
are not influenced by the mass of the external gas swept up by the
bow shock. The dynamical behaviour of active radio sources
strongly suggests that their cocoons are underdense with respect
to the external medium \citep[e.g.][]{sf91}. Therefore it is very
unlikely that this assumption can be made for the problem
considered here.

\subsection{Analytical solutions}

For simplicity we will assume that the bow shock and the cocoon are
both spherical. In practice both are elongated along the jet axis but
this different geometry results only in a change of the constants of
proportionality. The dynamical behaviour, i.e. the time dependence of
the physical parameters of the flow, will be the same. The cocoon
volume is then given by $V_{\rm c} = 4/3 \pi R_{\rm c}^3$ and that of
the layer of shocked gas in between the bow shock and the cocoon is
$V_{\rm s} = 4/3 \pi (R_{\rm s}^3-R_{\rm c}^3)$. Furthermore, we also
assume that the pressure, $p$, inside the cocoon and in the shocked
layer surrounding it, is uniform. Again, this greatly simplifies the
problem and will not change our results apart from the exact numerical
values of the normalisations. During the active phase, while the jets
are still switched on, conservation of energy for the gas inside
$V_{\rm c}$ requires

\begin{equation}
\frac{4}{3} \pi R_{\rm c}^3 \dot{p} + 4 \pi \gamma _{\rm c} p R_{\rm c}^2 \dot{R}_{\rm c} = \left( \gamma _{\rm c} -1 \right) Q_{\rm j},
\label{cocoon}
\end{equation}

\noindent where $\gamma _{\rm c}$ is the adiabatic index of the cocoon material and $Q_{\rm j}$ is the energy transport rate of the jets. We will keep $Q_{\rm j}$ constant during the lifetime of the jets at the end of which it drops to zero instantaneously. Similarly, for the shocked gas inside $V_{\rm s}$ we find

\begin{eqnarray}
\frac{4}{3} \pi \left( R_{\rm s}^3 - R_{\rm c}^3 \right) \dot{p} & + & 4 \pi \gamma _{\rm s} p \left( R_{\rm s}^2 \dot{R}_{\rm s} - R_{\rm c}^2 \dot{R}_{\rm c} \right) \nonumber\\
& = & 2 \pi \left( \gamma _{\rm s} -1 \right) R_{\rm s}^2 \rho _0 \left( \frac{R_{\rm s}}{a_0} \right) ^{-\beta} \dot{R}_{\rm s}^3,
\label{shock}
\end{eqnarray}

\noindent where $\gamma _{\rm s}$ is the adiabatic index of the shocked material. The expression on the right of equation (\ref{shock}) is the rate at which kinetic energy flows into the bow shock \citep[see][]{ka96b}. Finally, since we consider only strong shocks, the pressure within the source is balanced by the ram pressure of the external medium,

\begin{equation}
p = \rho _0 \left( \frac{R_{\rm s}}{a_0} \right) ^{-\beta} \dot{R}_{\rm s}^2.
\label{ram}
\end{equation}

\noindent Equations (\ref{cocoon}) through (\ref{ram}) are identical to the system of equations studied by \citet{rb97} for the case of intermittent jet sources.

We are interested in steady-state similarity solutions for this set of
equations with the three unknown functions $R_{\rm c}$, $R_{\rm s}$
and $p$ being power laws of time. We find

\begin{eqnarray}
R_{\rm c} & = & \left[ \frac{ \left( \gamma _{\rm c} -1 \right) \left( 5 -\beta \right)^3}{12 \pi \left(9 \gamma _{\rm c} -4 - \beta \right)} k^{\frac{\beta -2}{5 -\beta}} \right] ^{1/3} \left( \frac{ Q_{\rm j} t^3}{\rho _0 a_0^{\beta}} \right) ^{1/\left( 5 - \beta \right)} \nonumber\\
R_{\rm s} & = & \left( k \frac{Q_{\rm j}t^3}{\rho _0 a_0^{\beta}} \right) ^{1/\left( 5 -\beta \right)} \label{active}
\\
p & = & \frac{9}{\left( 5 - \beta \right) ^2} \left[ k ^{2-\beta} Q_{\rm j} ^{2-\beta} \left( \rho _0 a_0^{\beta} \right) ^3 t^{-4-\beta} \right]^{1/ \left( 5 - \beta \right)} \nonumber,
\end{eqnarray}

\noindent where

\begin{equation}
k = \frac{ \left( \gamma _{\rm c} -1 \right) \left( 5 -\beta \right) ^3 \left( 9 \gamma _{\rm s} -4-\beta \right)}{6 \pi \left( 9 \gamma _{\rm c} -4-\beta \right) \left( 9 \gamma _{\rm s} +1 -2\beta \right)}.
\end{equation}

The exponents of the time variable in these equations are equal to
those found by \citet{sf91} and \citet{ka96b} for sources with active
jets.

After the jets cease to supply energy to the cocoon, the source
enters the coasting phase and we have to set $Q_{\rm j} =0$. We
assume that the cocoon stays overpressured with respect to the
surrounding medium. In this case, equation (\ref{cocoon})
describes the now adiabatic expansion of the cocoon, i.e. $p
\propto R_{\rm c}^{-3 \gamma _{\rm c}}$. With this, equations
(\ref{shock}) and (\ref{ram}) in principle again yield power-law
similarity solutions with $R_{\rm c} \propto R_{\rm s} \propto
t^{2 / \left( 2 -\beta + 3 \gamma _{\rm c} \right)}$. Note that
this is different from the Taylor-Sedov solution. This difference
is caused by the adiabatic behaviour of the cocoon driving the bow
shock for which there is no analogon in the case of a strong
explosion. The ratio of the radius of the spherical cocoon and the
bow shock is given by

\begin{equation}
\frac{R_{\rm c}}{R_{\rm s}} = \left[ 1 - \frac{\gamma _{\rm s} -1}{2 \left( \gamma _{\rm s} - \gamma _{\rm c} \right)} \right] ^{1/3}.
\end{equation}

\noindent Obviously, this solution cannot be correct for $\gamma _{\rm c} = \gamma _{\rm s}$ or $\gamma _{\rm s}=2 \gamma _{\rm c} -1$ which includes the case of a relativistic cocoon ($\gamma _{\rm c} = 4/3$) surrounded by `cold' material ($\gamma _{\rm s} = 5/3$). Nevertheless, we will see in section \ref{sec:num} that in the latter scenario the steady-state solution is an excellent approximation to the evolution of $R_{\rm c}$.

For $\gamma _{\rm c} = \gamma _{\rm s}$ all terms containing
$R_{\rm c}$ in equation (\ref{shock}) cancel and a different
steady-state solution is possible with

\begin{eqnarray}
R_{\rm c} & \propto & t^{\frac{2 \left( \gamma _{\rm c} +1 \right)}{\gamma _{\rm c} \left( 7 + 3\gamma _{\rm c} - 2 \beta \right)}} \nonumber\\
R_{\rm s} & \propto & t^{\frac{4}{7 + 3\gamma _{\rm c} - 2 \beta}}\label{same}
\\
p & \propto & t^{\frac{-6 \left( \gamma _{\rm c} +1 \right)}{7 + 3\gamma _{\rm c} - 2 \beta}}. \nonumber
\end{eqnarray}

\noindent This implies that the ratio $R_{\rm c} / R_{\rm s}$ is no longer a constant. Despite this, the expansion of the bow shock and that of the cocoon are still intrinsically self-similar.

\subsection{Numerical solutions}
\label{sec:num}

All of the solutions derived in the previous section describe
steady-state configurations. To assess how quickly a spherical radio
source would change from one of these states to another when the jets
are switched off, we also integrated the equations (\ref{cocoon}) to
(\ref{ram}) numerically. For this we set $Q_{\rm j} =
10^{46}$\,ergs\,s$^{-1}$, $\rho _0 = 1.7 \times
10^{-25}$\,g\,cm$^{-3}$, $a_0 = 500$\,pc and $\beta =1.5$. The same
values were used by \citet{rb97} and they imply that $\rho _0
a_0^{\beta} = 10^7$\,g\,cm$^{-1.5}$. At a source lifetime of
$10^7$\,years, we `switched off' the jets by setting $Q_{\rm
j}=0$. For the integration we used a standard Runge-Kutta
algorithm. The initial conditions were derived from equations
(\ref{active}).

\begin{figure}
\centerline{
\includegraphics[width=8.45cm]{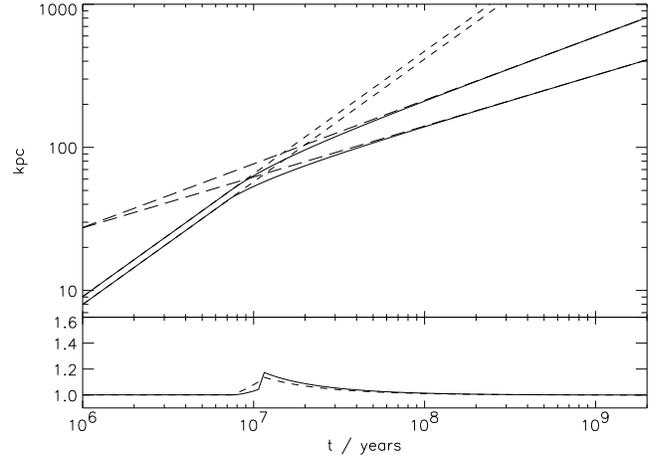}}
\caption{Comparison of the numerical results with the analytical solutions for $\gamma _{\rm c} = \gamma _{\rm s} =5/3$. Upper panel, solid lines: Numerical results for $R_{\rm c}$ (lower line) and $R_{\rm s}$ (upper line); short dashed lines: analytical solution for active jets; long dashed lines: analytical solutions for `coasting phase'. Lower panel: Analytical solutions divided by numerical solutions. Dashed line: Ratio for $R_{\rm c}$, solid line: Ratio for $R_{\rm s}$.}
\label{fig:same}
\end{figure}

Figure \ref{fig:same} shows the results of this numerical
integration for the radius of the bow shock, $R_{\rm s}$, and that
of the cocoon, $R_{\rm c}$ for the case $\gamma _{\rm c} = \gamma
_{\rm s} = 5/3$. The analytical solutions for the still active
jets are calculated from equations (\ref{active}). For the
coasting phase we used equations (\ref{same}). Here we are free to
choose the normalisation of the analytical solutions as these are
not constrained by the equations. We have chosen the normalisation
such that the analytical solution agrees with the numerical
results at $2\times 10^9$\,years. The lower panel of Figure
\ref{fig:same} shows the ratio of the analytical solutions and the
numerical results. We have used the analytical solutions for the
active phase for times less than $10^7$\,years and those for the
coasting phase for later times. The discrepancy between the two
solutions is always less than 20\%.

\begin{figure}
\centerline{
\includegraphics[width=8.45cm]{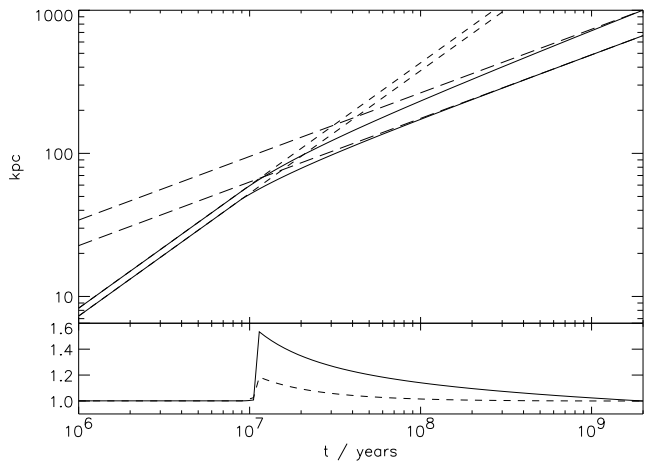}}
\caption{Same as figure \ref{fig:same} but with $\gamma _{\rm c} =4/3$ and $\gamma _{\rm s} = 5/3$.}
\label{fig:dif}
\end{figure}

Figure \ref{fig:dif} compares the numerical results for the case
$\gamma _{\rm c} = 4/3$ and $\gamma _{\rm s} = 5/3$ with the
steady-state similarity solutions. As expected, the agreement is
worse. Particularly the radius of the bow shock grows more quickly
than predicted by the power-law solution. However, the radius of
the cocoon is still well described by the steady-state solution.
The disagreement between the numerical integration and the
analytical power-law is comparable for this case to that for
$\gamma _{\rm c} = \gamma _{\rm s}$. We therefore conclude that
the dynamical behaviour of the cocoon of a spherical radio source
adjusts quickly to the cessation of the energy supply by the jets.

\subsection{Application to `real' radio sources}

The cocoons of radio sources are elongated along the jet axis. Because
of the different external density the various source parts are
immersed in, this deviation from purely spherical shape may lead to a
different dynamical behaviour at the tip of the cocoon compared to its
base. In steady-state this adds only numerical factors to the
solutions discussed above. However, it may prolong the transition to
the coasting phase at the end of the lifetime of the jets. To
investigate this possibility, fully 3-dimensional simulations of radio
sources making the transition from active to coasting are needed
\citep*[e.g.][]{rhb00}. To date the evidence from these simulations
is somewhat inconclusive. Nevertheless, it is unlikely that the
geometrical shape of the radio source strongly influences its
evolutionary behaviour.

So far we have implicitly assumed that the information that the jets
have ceased to supply energy to the cocoon propagates instantaneously
through the whole cocoon. This is a good approximation if the cocoon
is small and the internal sound speed is very high. \citet{ka96b} show
that in the absence of significant mixing of the cocoon material with
the surrounding gas across the cocoon boundary the internal sound
speed is indeed high. In cases where mixing is important, the sound
travel time along the cocoon may become comparable to or even exceed
the dynamical timescale. Most of the cocoon will then continue for a
significant time to expand as if it was still in the active phase
\citep*[e.g.][]{ksr00}.

In any scenario the pressure in the cocoon will eventually become
comparable to the thermal pressure of the surrounding gas. At this
point, the expansion will slow down and the bow shock weakens. Once
pressure equilibrium is reached, the expansion stops altogether and
the pressure does not change anymore. When exactly this happens
depends on the temperature of the external gas. Again the transition
of the cocoon from expansion to pressure equilibrium will be a gradual
process with those regions of the cocoon close to the centre of the
density distribution coming into pressure equilibrium first
\citep[e.g.][]{jb82}. Even in this equilibrium phase the cocoon will
not be static. As it is underdense with respect to the surrounding gas
it will start rising due to buoyancy \citep[e.g.][]{gn73,cbkbf00,bk01}
and the pressure inside the cocoon is decreasing. However, the
timescale for this rise usually exceeds the lifetime of the
relativistic electrons. The equilibrium phase was also considered by
\citet{kg94}. For simplicity we study in this paper the case of a
radio source which comes into pressure equilibrium with its
surroundings at the same time as the jets switch off. We will also
assume that the transition is instantaneous and that the pressure
within the cocoon is uniform.

In the following study of the radio emission from coasting or
equilibrium radio sources we will concentrate on three limiting
cases. In each case the jets stop to supply energy to the cocoon
at time $t_{\rm s} < t$, where $t$ is the time of observation.
\begin{enumerate}
\item[Model A:] The internal sound speed is slow and so the cocoon continues to evolve as if the jets were still supplying it with energy, i.e. $R_{\rm c} \propto t^{3/\left( 5-\beta \right)}$ and $p \propto  t^{\left( -4-\beta \right)/ \left( 5-\beta \right)}$ even for $t > t_{\rm s}$.
\item[Model B:] The internal sound speed is fast and so the cocoon immediately changes from the active phase to the coasting phase, i.e. for $\gamma _{\rm c} =\gamma _{\rm s}=5/3$ and $t > t_{\rm s}$ we use equations (\ref{same}).
\item[Model C:] The source is in pressure equilibrium with its surroundings from the time $t_{\rm s}$. In this case $R_{\rm c}$ and $p$ are both constants between $t_{\rm s}$ and $t$.
\end{enumerate}

\section{Radio emission from an inactive radio source}
\label{sec:radio}

The relativistic electrons responsible for the emission of radio
synchrotron emission are accelerated by the shocks at the end of
the active jets and injected into the cocoon. They subsequently
loose their energy due to adiabatic expansion of the cocoon,
synchrotron radiation and inverse Compton scattering of CMB
photons. All these processes are cumulative and so relativistic
electrons injected into the cocoon later may still radiate while
material injected earlier may have faded already. The model of
\citet*{kda97a} for active sources allows to track the evolution
of the energy distribution of the relativistic electrons injected
at a given time $t_{\rm i}$. The total radio spectrum is then
found by integrating up the contributions of all electrons
injected into the cocoon between $t_{\rm i} =0$ and $t_{\rm i} =
t$, where $t$ is the current age of the source. In the following
we will use this model with the modifications detailed in
\citet{ck00b} also for the calculation of the emission of sources
in the coasting phase.

The model employed relies on all relevant physical quantities
having power-law dependencies on time. We have seen in the
previous sections that this is, at least approximately, the case
for the coasting phase. We find the total radio emission of the
cocoon for Model A by simply setting the upper limit of the
integration over $t_{\rm i}$ to $t_{\rm s} < t$ at which the jets
switched off. In addition, for Models B and C we have to slightly
modify the basic approach of \citet{kda97a}. The relativistic
electrons injected into the cocoon at time $t_{\rm i}$ during a
short time interval $\delta t_{\rm i}$ are contained in a volume
$\delta V$. To determine the radio synchrotron emission from
$\delta V$, we need to know the energy distribution of the
relativistic electrons and the energy density of the magnetic
field inside $\delta V$. These are completely determined by the
evolution of $\delta V$ and the pressure, $p$, as functions of
time. Defining $\delta V_{\rm s}= \delta V \left( t_{\rm s}
\right)$ and $p_{\rm s} = p \left( t_{\rm s} \right)$, yields for
Model B at $t>t_{\rm s}$

\begin{equation}
\delta V = \delta V_{\rm s} \left( \frac{t}{t_{\rm s}} \right)
^{\frac{6\left( \gamma _{\rm c} +1 \right)}{\gamma _{\rm c} \left(
7 +3\gamma _{\rm c} -2\beta \right)}} \ {\rm and} \ p= p_{\rm s}
\left( \frac{t}{t_{\rm s}} \right) ^{\frac{6\left( \gamma _{\rm c}
+1 \right)}{7 +3\gamma _{\rm c} -2\beta}},
\end{equation}

\noindent where we have assumed that each volume element $\delta
V$ undergoes adiabatic expansion with $\delta V \propto R_{\rm
c}^3$. This follows naturally as the sum of all volume elements
$\delta V$ must be equal to the total volume of the cocoon. The
time dependencies are then given by equations (\ref{same}).
Obviously, for Model C we have $\delta V = \delta V_{\rm s}$ and
$p=p_{\rm s}$ for all times $t>t_{\rm s}$. With this it is now
straightforward to compute the synchrotron emission of each
$\delta V$ and numerically integrate over the injection time
$t_{\rm i}$ as described in \citet{kda97a}.

\citet{kda97a} assume efficient pitch-angle scattering for the
relativistic electrons moving in the magnetic field. This is
usually referred to as JP-type models \citep{jp73}. For
inefficient scattering, the electrons with large pitch-angles,
$\theta$, loose their energy faster than those with small $\theta$
\citep[KP-type models:][]{nk62,ap70}. It is straightforward to
modify the present model to account for negligible pitch-angle
scattering by assuming a uniform pitch-angle distribution and the
introduction of one further integral over $\theta$. We calculate
the radio emission from all models with and without pitch-angle
scattering.

The model for the radio emission depends on the following
parameters \citep[see also][]{kda97a,ck00b}: $t$ the current age
of the source, $t_{\rm s}$ the time at which the jets switched
off, $p_{\rm s}$, the pressure in the cocoon at $t_{\rm s}$,
$R_{\rm T}$, the aspect ratio of the cocoon, $s$, the exponent of
the initial power-law energy distribution of the relativistic
electrons, $\gamma _{\rm max}$, the high energy cut-off of this
distribution, $\beta$, the exponent of the density distribution of
the external medium and, $z$, the cosmological redshift of the
source. \citet{ck00b} shows that most of these parameters are
degenerate in the sense that their values do not determine the
shape of the radio spectrum in a unique way. We therefore
concentrate on the effects of changing $p_{\rm s}$, $t$ and
$t_{\rm s}$ and keep all other parameters fixed at their fiducial
values given in table \ref{tab:fiducial}. We also assume that, at
least at the time of injection into the cocoon, the magnetic field
is in equipartition with the relativistic particles in each
$\delta V$. We neglect any contribution to the cocoon pressure
from non-radiating particles, i.e. thermal electrons or protons.
With these assumptions and free parameters a model spectrum is
completely determined except for a multiplicative normalisation
constant, $A_{\rm f}$. For the fiducial model we set $A_{\rm f}
=1$. From $p_{\rm s}$, $t_{\rm s}$ and $A_{\rm f}$ it is possible
to infer the energy transport rate of the jets, $Q_{\rm j}$, and
$\rho _0 a_0^{\beta}$, which describes the density distribution of
the gas surrounding the source \citep[see][for details]{ck00b}.

\begin{table}
\caption{Model parameters for the fiducial model (see text).}
\begin{tabular}{lcc}
\hline & $R_{\rm T}$ & $2$\\ & $s$ & $2.5$\\ fixed parameters &
$\gamma _{\rm max}$ & $10^6$\\ & $\beta$ & $1.5$\\ & $z$ & 0.1\\
\hline & $t$ / years & $10^8$\\ free parameters & $t_{\rm s}$ /
years & $10^8$\\ & $p_{\rm s}$ / ergs cm$^{-3}$ & $3 \times
10^{-13}$\\ \hline & $Q_{\rm j}$ / ergs s$^{-1}$ & $10^{46}$\\
\raisebox{5pt}[5pt]{derived parameters} & $\rho _0 a_0^{\beta}$ /
g cm$^{-1.5}$ & $10^7$\\ \hline
\end{tabular}
\label{tab:fiducial}
\end{table}

\begin{figure}
\centerline{
\includegraphics[width=8.45cm]{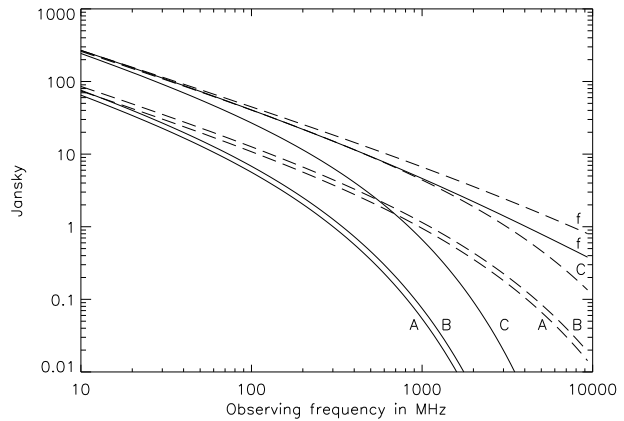}}
\caption{Model results for various coasting radio sources. Solid
lines: With efficient pitch-angle scattering, dashed lines:
Without pitch-angle scattering. Labels refer to the various models
defined in the text. `f' is the fiducial model and curves labeled
`A', `B' and `C' are for the same source but with $t=2\times
10^8$\,years.} \label{fig:models}
\end{figure}

For the fiducial model as described in table \ref{tab:fiducial},
$t=t_{\rm s}$ and so the jets are still active. The curves labeled
`f' in Figure \ref{fig:models} show the spectrum for this model
with (JP) and without (KP) pitch-angle scattering. With an age of
$10^8$\,years, the fiducial model predicts a fairly aged
synchrotron spectrum. However, the cumulative energy losses of the
relativistic electrons due to synchrotron radiation and inverse
Compton scattering are more severe at high frequency for the
JP-type model. The other model spectra in Figure \ref{fig:models}
are for the same model parameters as the fiducial model but with
an age $t=2\times 10^8$\,years which implies a coasting phase of
$10^8$\,years after $t_{\rm s}$. The spectral shape is very
similar among all JP-type and KP-type models so that they appear
to be the same spectrum simply shifted along the two coordinate
axes. This is caused by the, for the chosen parameters, rather
weak energy losses of the relativistic electrons after $t_{\rm
s}$. The break in the energy distribution of these electrons is
set mainly by the losses before $t_{\rm s}$ and so occurs at
roughly the same energy in all models. The break in the spectrum
resulting from this is then simply shifted in frequency/flux space
by the strength of the magnetic field. Since the energy density of
the magnetic field is proportional to the pressure in the cocoon,
it follows from Section \ref{sec:spherical} that models A and C
should have the lowest and highest break frequency respectively.
This is confirmed by Figure \ref{fig:models}. Note here that for
our choice of $\gamma _{\rm c}$ and $\beta$ the behaviour of $p$
as a function of time for models A ($p \propto t^{-11/7}$) and B
($p \propto t^{-16/9}$) is very similar, even for cases with
larger energy losses of the relativistic electrons. In fact,
usually $0\le \beta \le 2$ and as long as $\gamma _{\rm c} =5/3$,
Models A and B are always similar. As expected, the spectral break
occurs at higher frequencies for the KP-type models compared to
the JP-type models.

\begin{figure}
\centerline{
\includegraphics[width=8.45cm]{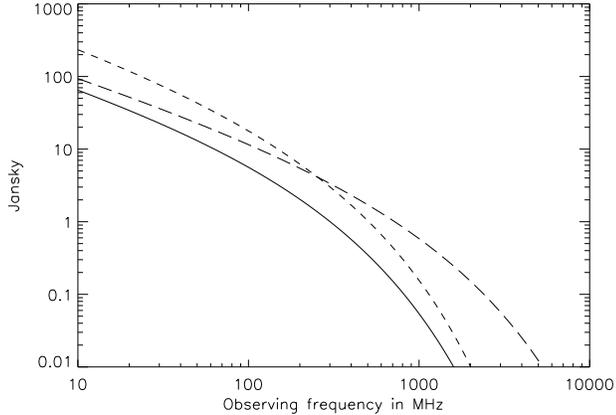}}
\caption{Model results for varying model parameters. Solid curve:
Model A for fiducial parameters but with $t=2\times 10^8$\,years
and efficient pitch-angle scattering (same as solid line labeled
`A' in Figure \ref{fig:models}). Short-dashed curve: Same model
but with $p_{\rm s} = 1.5 \times 10^{-12}$. Long-dashed curve:
Same model but with $t=2\times 10^7$\,years and $t_{\rm s} =
10^7$\,years.} \label{fig:paras}
\end{figure}

Figure \ref{fig:paras} shows a comparison of the effects of
varying the free parameters $t$, $t_{\rm s}$ and $p_{\rm s}$. For
this comparison we use the fiducial Model A with effective
pitch-angle scattering and $t=2\times 10^8$\,years. Since the
energy densities of the magnetic field and that of the
relativistic electrons is proportional to the pressure in the
cocoon, an increased $p_{\rm s}$ results in a higher radio flux at
low frequencies. However, the energy losses of the particles due
to synchrotron emission are also increased moving the break in the
spectrum to lower frequencies. Keeping the ratio $t_{\rm s}/t$
constant but decreasing the source age to $t=10^7$\,years, results
in a brighter source at low frequencies with the spectral break
moving to higher frequencies. At low frequencies the younger age
implies a higher cocoon pressure which leads to an increased radio
flux. As the radiative energy losses of the relativistic electrons
are cumulative, the younger source age simply translates to
smaller total losses and therefore a higher break frequency.

Finally we note here that the injection of the relativistic
electrons by the jets into the cocoon over a range of time leads
to a flatter spectrum at high frequencies than for a single burst
injection \citep[e.g.][]{jl91}. `Mild' breaks observed in the
spectra of radio sources are often interpreted, when using single
burst injection models \citep[as assumed in][]{kg94}, as resulting
from the suppression of pitch-angle scattering. For a range in the
injection time, a similar flattening may be achieved even for
efficient pitch-angle scattering.

\section{Comparison with the spectra of radio relics}
\label{sec:compa}

In this section we will use the models developed above to fit the
observed spectra of radio relic sources. In order to compare our
results with those of \citet{kg94}, we use their data points and error
estimates for the same relic sources, i.e. Mol 0100-22 (or MRC
0100-221), Cul 0038-096, 4C 38.39, 4C 63.10, 3C 318 and 3C 464. The
names of the clusters these sources are found in is given in table
\ref{tab:paras}. As discussed in \citet{kg94}, there is a
contradiction between the flux measurements of \citet{sr84} and
\citet*{sps89} at 1.465\,GHz for Cul 0038-096. We follow \citet{kg94}
in using both flux measurements separately, assigning indices `a' and
`b' to the different spectra. However, we note that \citet{bpl98} and
the relic flux density measure from the NVSS are in agreement with
\citet{sr84}. We also fit our models to the radio spectrum of the
relic source 1253+275. The relic is somewhat different to the six
others in that it is rather extended and shows practically no break in
its radio spectrum. Because of these differences, 1253+275 may have
had a completely different origin and/or history compared to the other
relics. However, we will show below that the radio spectrum of this
source is also well fitted by our models. The radio fluxes for
1253+275 at seven radio frequencies can be found in \citet{gfs91}

We fit the spectra adopting the fixed model parameters of the fiducial
model and using $t$, $t_{\rm s}$ and $p_{\rm s}$ as free
parameters. The goodness of the fit is determined by the standard
$\chi ^2$-technique. The normalisation of the spectrum, $A_{\rm f}$,
is chosen such that for each set of free parameters the $\chi
^2$-value is minimised. The minimum of $\chi ^2$ as a function of the
free parameters is found using a downhill simplex method in three
dimension \citep[e.g.][]{ptvf92}. To avoid solutions with excessively
young or old sources, we restricted $t$ to the interval from $5 \times
10^6$\,years to $5 \times 10^8$\,years.

\begin{figure}
\centerline{
\includegraphics[width=8.45cm]{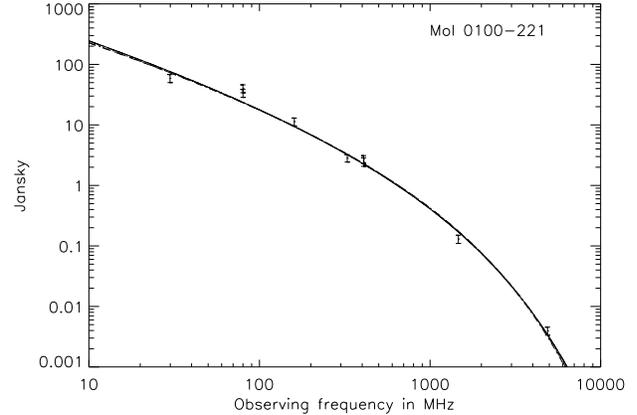}}
\caption{Model fits to the observed spectrum of Mol 0100-221.
Subscripts `JP' denote models with efficient pitch-angle
scattering, subscripts `KP' are assigned to models without
pitch-angle scattering. Flux measurements are shown with their
errors. Dotted line: Model A$_{\rm JP}$. Short-dashed line: Model
B$_{\rm JP}$. Dash-dotted line: Model C$_{\rm JP}$. Dash-triple
dotted line: Model A$_{\rm KP}$. Long-dashed line: Model B$_{\rm
KP}$. Solid line: Model C$_{\rm KP}$.} \label{fig:mol0100}
\end{figure}

\begin{figure}
\centerline{
\includegraphics[width=8.45cm]{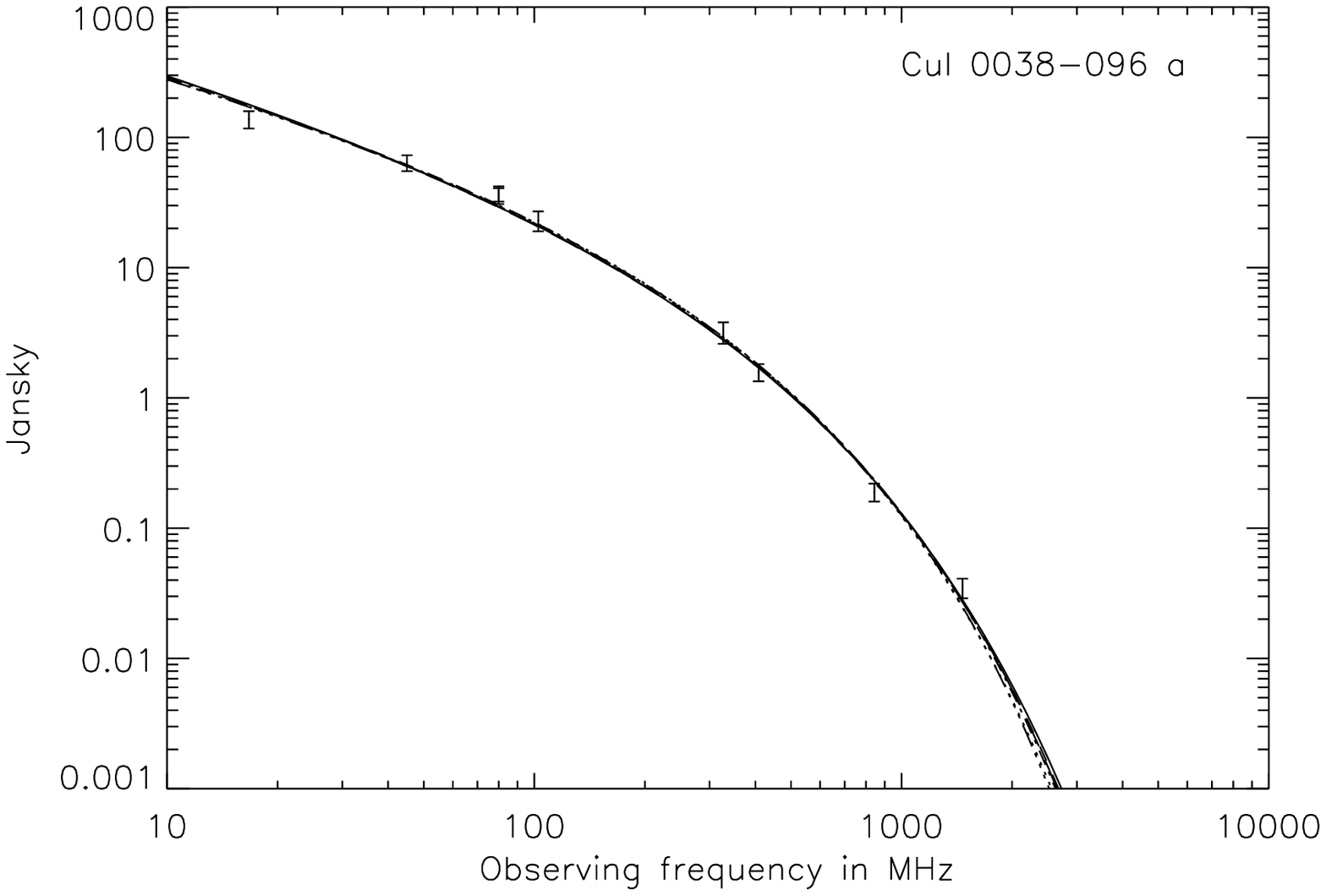}}
\caption{Same as Figure \ref{fig:mol0100} but for Cul 0038-096
using flux measurement of \citet{sr84} at 1.465\,GHz.}
\label{fig:cul0038a}
\end{figure}

\begin{figure}
\centerline{
\includegraphics[width=8.45cm]{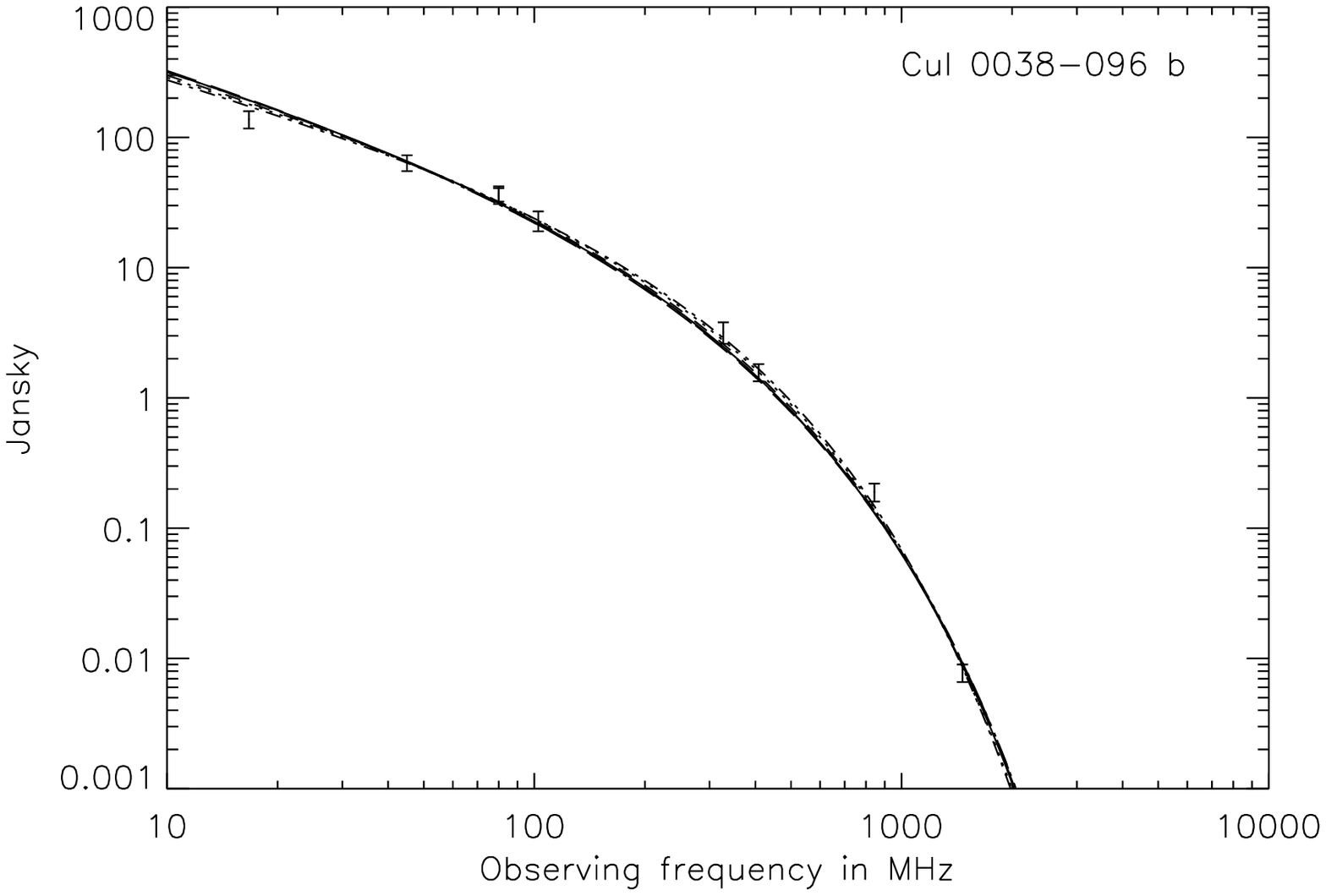}}
\caption{Same as Figure \ref{fig:mol0100} but for Cul 0038-096
using flux measurement of \citet{sps89} at 1.465\,GHz.}
\label{fig:cul0038b}
\end{figure}

\begin{figure}
\centerline{
\includegraphics[width=8.45cm]{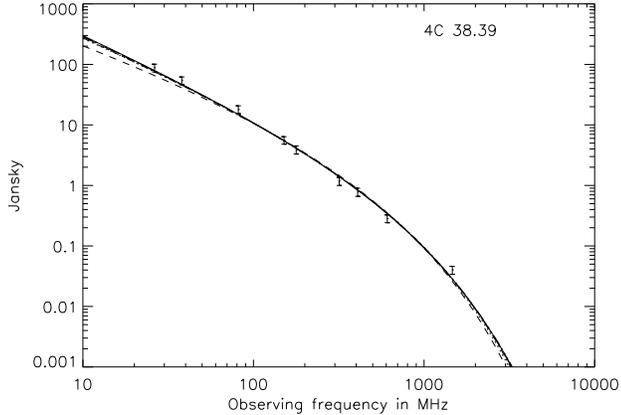}}
\caption{Same as Figure \ref{fig:mol0100} but for 4C 38.39.}
\label{fig:4c38}
\end{figure}

\begin{figure}
\centerline{
\includegraphics[width=8.45cm]{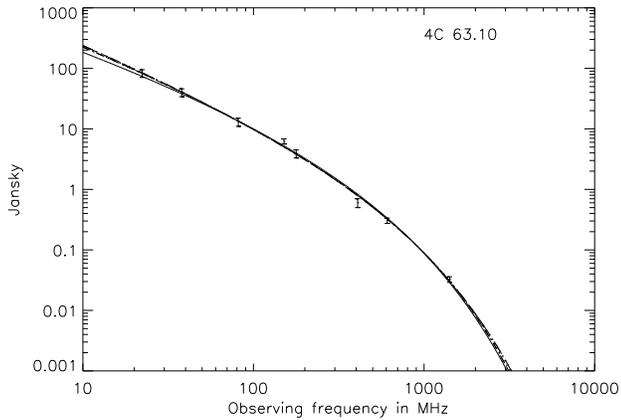}}
\caption{Same as Figure \ref{fig:mol0100} but for 4C 63.10.}
\label{fig:4c63}
\end{figure}

\begin{figure}
\centerline{
\includegraphics[width=8.45cm]{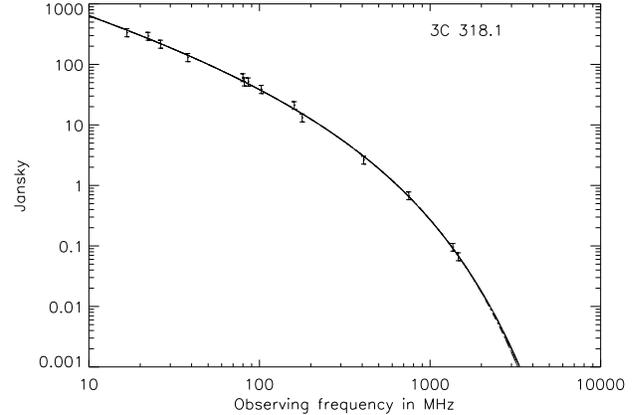}}
\caption{Same as Figure \ref{fig:mol0100} but for 3C 318.1.}
\label{fig:3c318}
\end{figure}

\begin{figure}
\centerline{
\includegraphics[width=8.45cm]{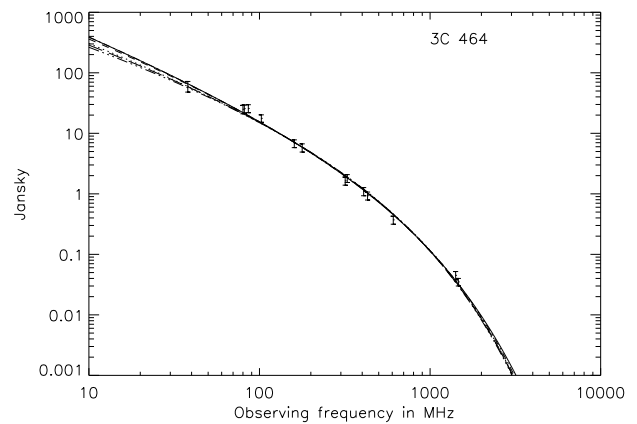}}
\caption{Same as Figure \ref{fig:mol0100} but for 3C 464.}
\label{fig:3c464}
\end{figure}

\begin{figure}
\centerline{
\includegraphics[width=8.45cm]{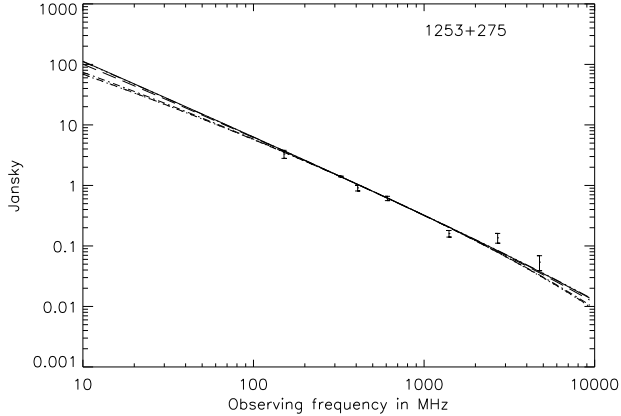}}
\caption{Same as Figure \ref{fig:mol0100} but for 1253+275.}
\label{fig:1253+275}
\end{figure}

\begin{table*}
\caption{Best fitting model parameters. The Abell number of the host
cluster of the respective radio relic is given below the source
name. Subscripts `JP' denote models with efficient pitch-angle
scattering, subscripts `KP' are assigned to models without pitch-angle
scattering.}
\begin{tabular}{clcccccc}
\hline & & $t$ / $10^6$\,years & $t_{\rm s}$ / $10^6$\,years & $p_{\rm
s}$ / $10^{-13}$\,ergs\,cm$^{-3}$ & $Q_{\rm j}$ /
$10^{44}$\,ergs\,s$^{-1}$ & $\rho _0 a_0^{\beta}$ /
$10^7$\,g\,cm$^{-1.5}$ & $\chi ^2$ / DOF\\ \hline & Model A$_{\rm JP}$
& $167$ & $139$ & $90.3$ & $2.6$ & $187$ & 6.13/6\\ & Model A$_{\rm
KP}$ & $41$ & $34$ & $2330$ & $0.71$ & $799$ & 5.91/6\\[2pt] Mol
0100-221 & Model B$_{\rm JP}$ & $176$ & $142$ & $67.9$ & $3.3$ & $133$
& 6.17/6\\ A133 & Model B$_{\rm KP}$ & $50$ & $42$ & $1654$ & $0.81$ &
$764$ & 5.92/6\\[2pt] & Model C$_{\rm JP}$ & $165$ & $136$ & $90.8$ &
$1.8$ & $193$ & 6.20/6\\ & Model C$_{\rm KP}$ & $37$ & $32$ & $3040$ &
$0.47$ & $993$ & 5.91/6\\ \hline & Model A$_{\rm JP}$ & $237$ & $102$
& $2.45$ & $140$ & $0.808$ & 2.80/6\\ & Model A$_{\rm KP}$ & $490$ &
$118$ & $11.1$ & $90$ & $6.69$ & 3.13/6\\[2pt] Cul 0038-096 a & Model
B$_{\rm JP}$ & $220$ & $123$ & $0.388$ & $440$ & $0.110$ & 2.49/6\\
A85 & Model B$_{\rm KP}$ & $185$ & $37$ & $92.3$ & $99$ & $9.26$ &
3.27/6\\[2pt] & Model C$_{\rm JP}$ & $171$ & $75$ & $34.9$ & $5.8$ &
$17.3$ & 3.13/6\\ & Model C$_{\rm KP}$ & $7.3$ & $3.1$ & $9160$ &
$1.2$ & $43.0$ & 3.07/6\\ \hline & Model A$_{\rm JP}$ & $153$ & $31$ &
$2.03$ & $1900$ & $0.0467$ & 2.26/6\\ & Model A$_{\rm KP}$ & $496$ &
$99$ & $13.9$ & $140$ & $5.84$ & 4.29/6\\[2pt] Cul 0038-096 b & Model
B$_{\rm JP}$ & $171$ & $41$ & $3.95$ & $840$ & $0.199$ & 3.12/6\\ A85
& Model B$_{\rm KP}$ & $499$ & $111$ & $13.1$ & $160$ & $6.63$ &
4.64/6\\[2pt] & Model C$_{\rm JP}$ & $259$ & $55$ & $0.874$ & $160$ &
$0.0766$ & 1.77/6\\ & Model C$_{\rm KP}$ & $500$ & $142$ & $21.9$ &
$5.14$ & $33.5$ & 3.95/6\\ \hline & Model A$_{\rm JP}$ & $5.3$ & $5.2$
& $2.40\times 10^5$ & $2.6$ & $4430$ & 3.05/6\\ & Model A$_{\rm KP}$ &
$130$ & $130$ & $9.84\times 10^4$ & $1.3$ & $6.51\times 10^5$ &
2.53/6\\[2pt] 4C 38.39 & Model B$_{\rm JP}$ & $440$ & $383$ & $44.7$ &
$17$ & $387$ & 4.67/6\\ A1914 & Model B$_{\rm KP}$ & $499$ & $494$ &
$4470$ & $2.0$ & $1.90\times 10^5$ & 2.67/6\\[2pt] & Model C$_{\rm
JP}$ & $27$ & $26$ & $5.18\times 10^4$ & $2.9$ & $1.4\times 10^4$ &
2.89/6\\ & Model C$_{\rm KP}$ & $5.1$ & $5.1$ & $2.91\times 10^7$ &
$0.61$ & $1.48\times 10^6$ & 2.48/6\\ \hline & Model A$_{\rm JP}$ &
$63$ & $61$ & $4920$ & $1.2$ & $5020$ & 2.18/5\\ & Model A$_{\rm KP}$
& $56$ & $55$ & $2.73\times 10^4$ & $0.44$ & $3.55\times 10^4$ &
2.08/5\\[2pt] 4C 63.10 & Model B$_{\rm JP}$ & $29$ & $28$ &
$1.14\times 10^4$ & $1.1$ & $3270$ & 2.20/5\\ A566 & Model B$_{\rm
KP}$ & $128$ & $124$ & $7740$ & $0.53$ & $3.59\times 10^4$ &
2.08/5\\[2pt] & Model C$_{\rm JP}$ & $5.2$ & $5.0$ & $1.07\times 10^5$
& $0.78$ & $2000$ & 2.22/5\\ & Model C$_{\rm KP}$ & $5.0$ & $4.3$ &
$7.49\times 10^4$ & $0.51$ & $1080$ & 2.80/5\\ \hline & Model A$_{\rm
JP}$ & $198$ & $141$ & $73.2$ & $6.2$ & $130$ & 1.37/11\\ & Model
A$_{\rm KP}$ & $106$ & $73$ & $533$ & $2.4$ & $459$ & 1.35/11\\[2pt]
3C 318.1 & Model B$_{\rm JP}$ & $203$ & $142$ & $64.6$ & $7.4$ & $110$
& 1.37/11\\ A2063 & Model B$_{\rm KP}$ & $27$ & $17$ & $2910$ & $2.6$
& $239$ & 1.34/11\\[2pt] & Model C$_{\rm JP}$ & $245$ & $186$ & $65.7$
& $3.0$ & $215$ & 1.36/11\\ & Model C$_{\rm KP}$ & $52$ & $38$ &
$1650$ & $0.91$ & $627$ & 1.34/11\\ \hline & Model A$_{\rm JP}$ &
$429$ & $382$ & $84.2$ & $1.8$ & $1180$ & 3.80/11\\ & Model A$_{\rm
KP}$ & $5.0$ & $3.8$ & $4.38\times 10^4$ & $0.62$ & $447$ &
4.48/11\\[2pt] 3C 464 & Model B$_{\rm JP}$ & $9.7$ & $9.2$ &
$4.06\times 10^4$ & $0.60$ & $2040$ & 3.27/11\\ A2626 & Model B$_{\rm
KP}$ & $5.0$ & $4.1$ & $6.36\times 10^4$ & $0.48$ & $830$ &
4.02/11\\[2pt] & Model C$_{\rm JP}$ & $25$ & $24$ & $2.87\times 10^4$
& $0.51$ & $8340$ & 2.99/11\\ & Model C$_{\rm KP}$ & $5.0$ & $4.8$ &
$5.64\times 10^5$ & $0.17$ & $1.68\times 10^4$ & 2.85/11\\ \hline &
Model A$_{\rm JP}$ & $488$ & $475$ & $31.3$ & $0.062$ & $1010$ &
3.57/4\\ & Model A$_{\rm KP}$ & $249$ & $248$ & $2.47\times 10^5$ &
$0.0074$ & $1.01\times 10^6$ & 3.25/4\\[2pt] 1253+275 & Model B$_{\rm
JP}$ & $493$ & $481$ & $33.0$ & $0.061$ & $1100$ & 3.56/4\\ Coma &
Model B$_{\rm KP}$ & $5.02$ & $5.01$ & $7.82\times 10^5$ & $0.0050$ &
$4.71 \times 10^4$ & 3.29/4\\[2pt] & Model C$_{\rm JP}$ & $498$ &
$489$ & $54.6$ & $0.048$ & $2040$ & 3.49/4\\ & Model C$_{\rm KP}$ &
$5.38$ & $5.38$ & $5.86\times 10^6$ & $0.0037$ & $5.90\times 10^5$ &
3.25/4\\ \hline
\end{tabular}
\label{tab:paras}
\end{table*}

The best-fitting model parameters are summarised in table
\ref{tab:paras} and the resulting spectra are plotted in figures
\ref{fig:mol0100} through \ref{fig:3c464}. In practically all
cases the best-fitting spectra from the various models are almost
indistinguishable from each other. In the following we briefly
discuss the results for each source individually.

{\em Mol 0100-221}\/: The spectrum is well fitted by all models
(reduced $\chi ^2$ between 0.99 and 1.03). The JP and KP-type models
give somewhat different results for the free model
parameters. However, within each family of models they are roughly
consistent with each other. This indicates that the shape of the
spectrum is mainly determined during the active phase of the
source. The exact behaviour of the cocoon in the subsequent coasting
phase has little influence. The JP-type models prefer older ages,
higher jet powers and less dense environments compared to the KP-type
models. KP-type models are marginally preferred over JP-type models.

{\em Cul 0038-096}\/: Both variants of the source spectrum can be
successfully fitted by the models. However, the goodness of fit and
the values of the free parameters vary significantly for each
variant. The best fit achieved has a reduced $\chi ^2$-value of only
0.3, but predicts a very low density of the external gas. Three of the
bets-fitting models (A$_{\rm KP}$ for variant a, A$_{\rm KP}$ and
B$_{\rm KP}$ for variant b) have ages very close to the maximum age,
$t= 5\times 10^8$\,years, allowed for the fit. At least for models A
and B the JP-type and KP-type models give similar results, but models
C are quite different from these. The spectrum of this source can be
fitted reasonably well with all the models discussed here. However, a
clear trend for the model parameters is not apparent.

{\em 4C 38.39}\/: For this source models with high external densities
are clearly preferred. However, the spread in the value of $\rho _0
a_0^{\beta}$ is considerable. There is also no agreement between the
models with respect to the source age. From figure \ref{fig:4c38} it
seems that the rather flat spectral break at high frequencies causes
problems for all models. Again there is no discernible trend in the
best-fitting parameters.

{\em 4C 63.10}\/: The spectrum of this source is fitted about
equally well by all models. In terms of the jet power, $Q_{\rm
j}$, and the external density parameter, $\rho _0 a_0^{\beta}$, JP
and KP-type models agree well with each other for models A and B,
but there is a large range for the predicted source age. Models C
require a somewhat lower density and they predict very low source
ages.

{\em 3C 318.1}\/: The model fits to the spectrum are very good
with a reduced $\chi ^2$ of around 0.12. Values for $Q_{\rm j}$
and $\rho _0 a_0^{\beta}$ are similar for all models but again for
this source there is a large spread for the source age.

{\em 3C 464}\/: Except for model A$_{\rm JP}$, all models predict
rather low source ages. The best fits are achieved by both models
C with rather high external densities. The predicted values of
$Q_{\rm j}$ are fairly uniform but the external density vary
strongly between models.

{\em 1253+275}\/: The model parameters agree with each other for
KP-type and JP-type models. However, the parameters are inconsistent
between the two groups. The KP-type models prefer a young source age
with virtually no influence of the phase in the source evolution when
the jets are switched off. The predicted densities and pressures are
very high. In the absence of radiative energy losses,our choice of
$p=2.5$ for the power law exponent of the energy distribution of the
relativistic electrons at injection time, implies a spectral index of
about $-0.75$. The observed spectrum of 1253+275 shows a spectral
index of around $-1.1$ with no spectral break. The in general flatter
spectra resulting from the KP-type models do not fit this steep
spectrum well. Therefore, the model requires high external densities
causing high pressure inside the lobes which subsequently show
substantial energy losses and steeper spectra. The predictions of the
JP-type model for an old source with moderate pressure values are more
realistic. The low jet powers predicted by all models are caused by
the substantially lower radio flux of this relic compared to the other
sources.

\section{Conclusions}
\label{sec:conc}

We have developed a model for the evolution of the lobes of radio
galaxies and radio-loud quasars of type FRII after the jets of the
central AGN have stopped supplying them with energy. In the case
that the lobes are still overpressured with respect to their
surroundings, we show that their volume is proportional to

\begin{equation}
V \propto \left\{
\begin{array}{ll}
t^{\frac{6 \left( \gamma _{\rm c}+1 \right)}{\gamma _{\rm c}
\left( 7 + 3 \gamma _{\rm c} -2 \beta \right)}} & {\rm \ ; for \ }
\gamma _{\rm c} = \gamma _{\rm s},\\

t^{\frac{6}{2 - \beta + 3 \gamma _{\rm c}}} & {\rm \ ; for \ }
\gamma _{\rm c} = 4/3 {\rm \ and \ } \gamma _{\rm s} = 5/3.\\
\end{array}
\right.
\end{equation}

\noindent The proportionality given above for $\gamma _{\rm c} =
\gamma _{\rm s}$ is, strictly speaking, not a solution of the
equations governing the hydrodynamics. However, we have shown that
is provides an excellent fit for the numerically found correct
solution. The numerical calculations also show that the transition
from the active phase with jets to the coasting phase without jets
is fast.

We identify the lobes in the coasting phase with radio relic
sources. The calculated model spectra depend only on a small
number of source and environment parameters: The total age of the
source, $t$, the age of the source when the jets were switched
off, $t_{\rm s}$, and the pressure inside the lobes at that time,
$p_{\rm s}$. We consider three limiting cases for the lobe
evolution: The lobes continue to expand as if the jets were still
supplying energy (Model A), the lobes continue to expand but with
the modified dynamics found for the coasting phase and summarised
above (Model B) and the lobes come into pressure equilibrium at
the time the jets switch off and stop expanding (Model C). We also
investigate the effects of efficient pitch-angle scattering
(JP-type models) and its absence (KP-type models).

All our models can provide satisfactory to excellent fits to the
observed spectra of radio relics. Taking into account the evolution of
the strength of the magnetic field inside the lobe during the
injection of relativistic electrons flattens the spectra sufficiently
to explain the observations. Thus we cannot rule out efficient
pitch-angle scattering as suggested by \citet{kg94}. Unfortunately,
the good fits provided by all models also imply that the radio spectra
of relic sources alone do not constrain any of the important source or
environment parameters.  The situation is aggravated by the findings
of \citet{srm01} that the spectra can also be fitted using the
alternative assumption of inhomogeneous magnetic fields inside the
relics. Furthermore, even the inactive lobes may be re-ignited by
compression during cluster mergers \citep{eg01}. Note here that the
model of \citet{eg01} for radio relics does not require the radio
plasma to originate in a powerful FRII-type radio galaxy.

The morphologies of radio relics are very varied and only rarely
resemble the well-defined lobes of active radio galaxies. This
strongly suggests that fluid flows in the clusters significantly
influence the evolution of radio galaxy remnants. Any re-acceleration
or energisation due to compression would make it even more difficult
to determine the original conditions in the relic. A possible solution
could be the combination of radio spectra with X-ray observations of
sufficient spatial resolution to identify inverse Compton scattered
photons originating in the relativistic plasma of the relics. The
X-ray observations could then be used to estimate the electron density
while the radio emission would constrain the strength of the magnetic
field. Despite the problems, it is interesting to note that our simple
models can fit the radio spectra of all relic sources studied
here. This includes the clearly curved spectrum of Cul 0038-096, a
relic with substantial substructure, as well as the spectrum of
1253+275 without any significant spectral break and a comparatively
smooth appearance. This would imply that, whatever the history of the
relic source, even a rather simplistic model can explain the observed
spectra in a variety of different ways thus making it impossible to
constrain the conditions within the relic sources. We cannot rule out
a complex history or re-acceleration of relativistic electrons
influencing the properties of the relic sources. However, in the light
of our results, such more complicated models are not required to fit
the available radio data.

The degeneracy of model parameters can, however, also be turned to
advantage in the case of statistical studies of the radio source
population. For example, we can use the known radio luminosity
function of active FRII objects and make predictions for the
cosmological distribution of their remnants. Because of the
mentioned degeneracy, these predictions will be almost
model-independent. Such statistical studies can be used to predict
the detection rate of radio galaxy remnants in low frequency radio
surveys with high surface brightness sensitivity (Cotter \&
Kaiser, in preparation).

In fact, it is not yet clear whether radio relic sources are
actually the remains of the lobes of powerful radio galaxies. Deep
observations of radio galaxies in which the jet flows have stopped
comparatively recently are necessary to decide whether the
conditions inside their lobes point towards them evolving into
relics \citep[e.g.][]{ksr00}. Clearly, the radio relic sources
continue to challenge our understanding of the physics of radio
galaxies and clusters.

\section*{Acknowledgments}

We would like to thank the anonymous referee for many helpful
suggestions.


\begin{thebibliography}{}

\bibitem[\protect\citeauthoryear{Bagchi, Pislar \& {Lima Neto}}{Bagchi
  et~al.}{1998}]{bpl98}
Bagchi J.,  Pislar V.,    {Lima Neto} G.~B.,  1998, MNRAS, 296, L23

\bibitem[\protect\citeauthoryear{Baldwin}{Baldwin}{1982}]{jb82}
Baldwin J.~E.,  1982, in Heeschen D.~S.,  Wade C.~M.,  eds, Extragalactic radio
  sources Evolutionary tracks of extended radio sources.
Reidel, Dordrecht, p.~21

\bibitem[\protect\citeauthoryear{Br{\"u}ggen \& Kaiser}{Br{\"u}ggen \&
  Kaiser}{2001}]{bk01}
Br{\"u}ggen M.,  Kaiser C.~R.,  2001, MNRAS, 325, 676

\bibitem[\protect\citeauthoryear{Churazov, Br{\" u}ggen, Kaiser, B{\" o}hringer
  \& Forman}{Churazov et~al.}{2001}]{cbkbf00}
Churazov E.,  Br{\" u}ggen M.,  Kaiser C.~R.,  B{\" o}hringer H.,    Forman W.,
   2001, ApJ, 554, 261

\bibitem[\protect\citeauthoryear{Eilek, Melrose \& Walker}{Eilek
  et~al.}{1997}]{emw97}
Eilek J.~A.,  Melrose D.~B.,    Walker M.~A.,  1997, ApJ, 483, 282

\bibitem[\protect\citeauthoryear{En{\ss}lin \& {Gopal-Krishna}}{En{\ss}lin \&
  {Gopal-Krishna}}{2001}]{eg01}
En{\ss}lin T.~A.,  {Gopal-Krishna} 2001, A\&A, 366, 26

\bibitem[\protect\citeauthoryear{Falle}{Falle}{1991}]{sf91}
Falle S. A. E.~G.,  1991, MNRAS, 250, 581

\bibitem[\protect\citeauthoryear{Fanaroff \& Riley}{Fanaroff \&
  Riley}{1974}]{fr74}
Fanaroff B.~L.,  Riley J.~M.,  1974, MNRAS, 167, 31

\bibitem[\protect\citeauthoryear{Giovannini \& Feretti}{Giovannini \&
  Feretti}{2000}]{gf00}
Giovannini G.,  Feretti L.,  2000, New Astron., 5, 335

\bibitem[\protect\citeauthoryear{Giovannini, Feretti \&
  Stanghellini}{Giovannini et~al.}{1991}]{gfs91}
Giovannini G.,  Feretti L.,    Stanghellini C.,  1991, A\&A, 252, 528

\bibitem[\protect\citeauthoryear{Giovannini, Tordi \& Feretti}{Giovannini
  et~al.}{1999}]{gtf99}
Giovannini G.,  Tordi M.,    Feretti L.,  1999, New Astron., 4, 141

\bibitem[\protect\citeauthoryear{Goldshmidt \& Rephaeli}{Goldshmidt \&
  Rephaeli}{1994}]{gr94}
Goldshmidt O.,  Rephaeli Y.,  1994, ApJ, 431, 586

\bibitem[\protect\citeauthoryear{Govoni, Feretti, Giovannini, B{\"o}hringer,
  Reiprich \& Murgia}{Govoni et~al.}{2001}]{gfg01}
Govoni F.,  Feretti L.,  Giovannini G.,  B{\"o}hringer H.,  Reiprich T.~H.,
  Murgia M.,  2001, A\&A, 376, 803

\bibitem[\protect\citeauthoryear{Gull \& Northover}{Gull \&
  Northover}{1973}]{gn73}
Gull S.~F.,  Northover K. J.~E.,  1973, Nat., 224, 80

\bibitem[\protect\citeauthoryear{Jaffe \& Perola}{Jaffe \& Perola}{1973}]{jp73}
Jaffe W.~J.,  Perola G.~C.,  1973, A\&A, 26, 423

\bibitem[\protect\citeauthoryear{Kaiser}{Kaiser}{2000}]{ck00b}
Kaiser C.~R.,  2000, A\&A, 362, 447

\bibitem[\protect\citeauthoryear{Kaiser \& Alexander}{Kaiser \&
  Alexander}{1997}]{ka96b}
Kaiser C.~R.,  Alexander P.,  1997, MNRAS, 286, 215

\bibitem[\protect\citeauthoryear{Kaiser, Dennett-Thorpe \& Alexander}{Kaiser
  et~al.}{1997}]{kda97a}
Kaiser C.~R.,  Dennett-Thorpe J.,    Alexander P.,  1997, MNRAS, 292, 723

\bibitem[\protect\citeauthoryear{Kaiser, Schoenmakers \& R{\"o}ttgering}{Kaiser
  et~al.}{2000}]{ksr00}
Kaiser C.~R.,  Schoenmakers A.~P.,    R{\"o}ttgering H. J.~A.,  2000, MNRAS,
  315, 381

\bibitem[\protect\citeauthoryear{Kardashev}{Kardashev}{1962}]{nk62}
Kardashev N.~S.,  1962, SvA, 6, 317

\bibitem[\protect\citeauthoryear{Kempner \& Sarazin}{Kempner \&
  Sarazin}{2001}]{ks01}
Kempner J.~C.,  Sarazin C.~L.,  2001, ApJ, 548, 639

\bibitem[\protect\citeauthoryear{Komissarov \& Gubanov}{Komissarov \&
  Gubanov}{1994}]{kg94}
Komissarov S.~S.,  Gubanov A.~G.,  1994, A\&A, 285, 27

\bibitem[\protect\citeauthoryear{Landau \& Lifshitz}{Landau \&
  Lifshitz}{1987}]{ll87}
Landau L.~D.,  Lifshitz E.~M.,  1987, Fluid mechanics, 2$^{\rm nd}$ edition.
Butterworth-Heinemann, Oxford

\bibitem[\protect\citeauthoryear{Leahy}{Leahy}{1991}]{jl91}
Leahy J.~P.,  1991, in Hughes P.~A.,  ed., Beams and jets in astrophysics
  Interpretation of large scale extragalactic jets.
Cambridge University Press, p.~100

\bibitem[\protect\citeauthoryear{Pacholczyk}{Pacholczyk}{1970}]{ap70}
Pacholczyk A.~G.,  1970, Radio Astrophysics.
Freeman, San Francisco

\bibitem[\protect\citeauthoryear{Press, Teukolsky, Vetterling \&
  Flannery}{Press et~al.}{1992}]{ptvf92}
Press W.~H.,  Teukolsky S.~A.,  Vetterling W.~T.,    Flannery B.~P.,  1992,
  Numerical Recipes. Second edition..
Cambridge University Press, Cambridge, UK.

\bibitem[\protect\citeauthoryear{Reynolds \& Begelman}{Reynolds \&
  Begelman}{1997}]{rb97}
Reynolds C.~S.,  Begelman M.~C.,  1997, ApJ, 487, L135

\bibitem[\protect\citeauthoryear{Reynolds, Heinz \& Begelman}{Reynolds
  et~al.}{2001}]{rhb00}
Reynolds C.~S.,  Heinz S.,    Begelman M.~C.,  2001, ApJ, 549, L179

\bibitem[\protect\citeauthoryear{Sedov}{Sedov}{1959}]{ls59}
Sedov L.~I.,  1959, Similarity and dimensional methods in mechanics..
Academic Press, London

\bibitem[\protect\citeauthoryear{Slee, Perley \& Siegman}{Slee
  et~al.}{1989}]{sps89}
Slee O.~B.,  Perley R.~A.,    Siegman B.~C.,  1989, Austral. J. Phys., 42, 633

\bibitem[\protect\citeauthoryear{Slee \& Reynolds}{Slee \&
  Reynolds}{1984}]{sr84}
Slee O.~B.,  Reynolds J.~E.,  1984, Proc. Astron. Soc. Aust., 5, 516

\bibitem[\protect\citeauthoryear{Slee, Roy, Murgia, Andernach \& Ehle}{Slee
  et~al.}{2001}]{srm01}
Slee O.~B.,  Roy A.~L.,  Murgia M.,  Andernach H.,    Ehle M.,  2001, AJ, 122,
  1172

\bibitem[\protect\citeauthoryear{Tribble}{Tribble}{1994}]{pt93}
Tribble P.~C.,  1994, MNRAS, 261, 57

\end{thebibliography}
\end{document}